\input harvmac.tex
%\draftmode
\baselineskip 16pt

\def\cn#1{{\hbox{ch}\left({#1}\right)}}
\def\ch#1#2{{\hbox{ch}_{#1}\left({#2}\right)}}
\def\td#1{{\hbox{Td}\left({#1}\right)}}
\def\co{{\cal O}}
\def\en{{\hbox{End}}}

\def\cl{{\cal L}}
\def\cF{{\cal F}}

\def\BZ{{\bf Z}}
\def\BP{{\bf P}}

\def\half{{1\over 2}}

\def\ket#1{|#1\rangle}

\lref\FW{D.S. Freed and E. Witten, ``Anomalies in string Theory with 
D-Branes'', hep-th/9907189.}
\lref\MM{R. Minasian and G. Moore, `` K-theory and 
Ramond-Ramond Charges'',
JHEP {\bf 11} (1997) 002, hep-th/9710230.}
\lref\bdlr{I. Brunner, M.R. Douglas, A. Lawrence and C. R\"omelsberger,
``D-branes on the quintic,'' hep-th/9906200.}
\lref\canda{P. Candelas, X. De La Ossa, A. Font, S. Katz and D.R. 
Morrison,``Mirror symmetry for two parameter models. 1,''
Nucl. Phys. {\bf B416}, 481 (1994) hep-th/9308083.}
\lref\candb{P.~Candelas, A.~Font, S.~Katz and D.R.~Morrison,
``Mirror symmetry for two parameter models. 2,''
Nucl. Phys. {\bf B429}, 626 (1994) hep-th/9403187.}
\lref\greene{B.R. Greene, ``String theory on Calabi-Yau manifolds,''
hep-th/9702155.}
\lref\grpl{B.R. Greene and M.R. Plesser,``Duality In Calabi-Yau Moduli
Space,'' Nucl.\ Phys.\ {\bf B338}, 15 (1990).}
\lref\stro{A. Strominger,``Massless black holes and conifolds in string
theory,'' Nucl.\ Phys.\ {\bf B451}, 96 (1995) hep-th/9504090.}
\lref\wiA{E.~Witten,``Phases of N = 2 theories in two dimensions,''
Nucl.\ Phys.\ {\bf B403}, 159 (1993) hep-th/9301042.}
\lref\wiB{E.~Witten, ``On the Landau-Ginzburg description of N=2 minimal
models,'' Int.\ J.\ Mod.\ Phys.\ {\bf A9}, 4783 (1994) hep-th/9304026.}
\lref\lvw{W.~Lerche, C.~Vafa and N.P.~Warner, ``Chiral Rings In N=2
Superconformal Theories,'' Nucl.\ Phys.\ {\bf B324}, 427 (1989).}
\lref\reso{A.~Recknagel and V.~Schomerus,``D-branes in Gepner models,''
Nucl.\ Phys.\ {\bf B531}, 185 (1998) hep-th/9712186.}
\lref\cardy{J.L.~Cardy, ``Boundary Conditions, Fusion Rules And The
Verlinde Formula,'' Nucl.\ Phys.\ {\bf B324}, 581 (1989).}
\lref\gep{D.~Gepner, ``Space-Time Supersymmetry In Compactified String
Theory And Superconformal Models,'' Nucl.\ Phys.\ {\bf B296}, 757 (1988).}
\lref\dofi{M.R.~Douglas and B.~Fiol, ``D-branes and discrete torsion. 
II'', hep-th/9903031.}

%%%%%%%%%%%%%%%%%%%%%%%%%%%%%%%%%%%%%%%%%%%%%%
\Title{\vbox{\baselineskip12pt\hbox{hep-th/9910172}
\hbox{IASSNS-HEP-99/97}
\hbox{RUNHETC-99-39}}}
{\vbox{\vskip-100pt
\hbox{\centerline{D-Branes and Bundles on Elliptic Fibrations}}}}
\vskip 10pt
\centerline{Duiliu-Emanuel Diaconescu$^\natural$ and 
Christian R\"omelsberger$^\sharp$}
\bigskip
\centerline{\it $^\natural$ School of Natural Sciences}
\centerline{\it Institute for Advanced Study}
\centerline{\it Olden Lane, Princeton, NJ 08540}
\centerline{\tt diacones@sns.ias.edu}
\medskip
\centerline{\it $^\sharp$ Department of Physics and Astronomy}
\centerline{\it Rutgers University}
\centerline{\it Piscatway, NJ 08555}
\centerline{\tt roemel@physics.rutgers.edu}
\bigskip
\bigskip
\bigskip
\noindent
We study the D-brane spectrum on a two-parameter Calabi-Yau model. 
The analysis is based on different tools 
in distinct regions of the moduli space: wrapped brane configurations 
on elliptic fibrations 
near the large radius limit, and SCFT boundary states at the 
Gepner point. We develop an explicit correspondence between these two 
classes of objects, suggesting that boundary states are natural quantum 
generalizations of bundles. We also find interesting D-brane 
dynamics in deep stringy regimes. The most striking example is, perhaps, 
that nonsupersymmetric D6-D0 and D4-D2 large radius configurations 
become stable BPS states at the Gepner point. 

\Date{October 1999}

\newsec{Introduction}

Since their discovery
\nref\branes{J. Dai, R. Leigh, and
J. Polchinski, "New Connections Between String Theories", Mod.
Phys. Lett. {\bf A4} (1989) 2073;\parskip=0pt
\item{}R. Leigh, "Dirac-Born-Infeld Action from Dirichlet Sigma Model",
Mod. Phys. Lett. {\bf A4} (1989) 2767;
\parskip=0pt
\item{}
P. Horava, ``Strings on World-Sheet Orbifolds,'' 
Nucl. Phys. {\bf B327} (1989) 461;
\parskip=0pt
\item{}
P. Horava, "Background Duality of Open String Models", Phys. Lett.
{\bf B231} (1989) 251;\parskip=0pt
\item{}
M.B. Green, "Space-time Duality and Dirichlet String Theory", Phys.
Lett. {\bf B266} 325 (1991),
"Pointlike States for  Type IIB Superstrings",
Phys. Lett. {\bf B329} (1994) 435, hep-th/9403040;
 "A Gas of D Instantons", Phys.
Lett. {\bf B354} (1995) 271, hep-th/9504108;\parskip=0pt
\item{}
J. Polchinski, "Combinatorics of Boundaries in String Theory",
Phys. Rev. {\bf D50} (1994) 6041.}%
\branes,\ D-branes have been an ubiquitous presence in string dualities 
and M theory. However, in spite of a detailed understanding of D-brane
dynamics in flat space, their behavior in abstract conformal field 
theories is less understood. 
A particular class 
of string vacua where D-brane spectra are especially interesting 
consists of $(2,2)$ superconfomal field theories (SCFT).
The moduli spaces of $(2,2)$ SCFT's are usually affected by stringy 
quantum corrections which result in a very rich phase structure.
One typically has geometric phases, where classical geometry 
can be used at least as a guiding principle in describing the physics, 
and nongeometric phases where the semiclassical description breaks 
down. It is known that the closed string quantum corrections result 
in a quantum deformation of the classical cohomology rings of 
varieties. Loosely, one can think of chiral primary operators 
in abstract SCFT's as representing quantum deformations of classical 
cohomology cycles. 

Adding D-branes is essentially equivalent 
to adding boundaries (with appropriate boundary conditions) to 
the string worldsheet. The coupling to the open string sector 
adds new ingredients to the space-time physics. In geometric phases,
the new degrees of freedom can be described semiclassically as gauge 
fields living on various submanifolds of space-time. Therefore, as 
explained in 
\nref\MM{R. Minasian and G. Moore, ``K-theory and Ramond-Ramond charge'',
JHEP {\bf 11} (1997) 002, hep-th/9710230.}%
\nref\W{E. Witten, ``D-Branes and K Theory'', JHEP
{\bf 12} (1998) 019, hep-th/9810188.}%
\refs{\MM,\W},
in a geometric phase, D-branes are naturally described as K theory 
classes rather than singular cohomology classes. 
However, such an
explicit and intuitive description is lacking in deep stringy regimes. 
Given the breakdown of the classical geometry in these regions, 
one has to rely on abstract SCFT techniques (whenever possible) 
in order to classify D-brane charges and study their dynamics.
An effective approach to this problem relies on the formalism of 
boundary states, which can be loosely thought as closed string coherent
states solving the SCFT boundary conditions. In this context, 
boundary states seem to be the natural quantum deformations of 
vector bundles. 

The main point of the present paper is the interplay between 
the two different 
descriptions of D-branes in $N=2$ string vacua.
The natural framework, which insures an explicit description in 
both regimes, consists of Calabi-Yau compactifications 
continuously connected to Gepner models. Since the latter are exactly 
solvable SCFT's one can construct explicit boundary states solving 
Cardy's consistency condition
\nref\reso{A. Recknagel and V. Schomerus,``D-branes in Gepner models,''
Nucl.\ Phys.\ {\bf B531}, 185 (1998) hep-th/9712186.}%
\nref\GSa{M. Gutperle and Y. Satoh, 
``D-branes in Gepner models and supersymmetry'', Nucl. Phys. 
{\bf B543} (1999) 73, hep-th/9808080.}%
\nref\GSb{M. Gutperle and Y. Satoh, ``D0-branes in Gepner models and 
N=2 black holes'', Nucl. Phys. {\bf B555} (1999) 477, hep-th/9902120.}%
\refs{\reso,\GSa,\GSb}.
This offers detailed information on part of the spectrum of branes in a 
nongeometric phase. As outlined above, we would like to understand 
if these states have a definite geometric interpretation in terms of 
D-branes wrapping supersymmetric cycles in a Calabi-Yau manifold 
of very large radius. Restricting our attention to the charges of 
BPS states (i.e. ignoring dynamical aspects such as stability and 
existence of bound states) this question can be systematically 
answered once 
the exact special geometry of the moduli space is known. The map
between the symplectic charge lattices can be found using the 
boundary state
representation of the symplectic intersection form described in 
\ref\DF{M.R. Douglas and B. Fiol, "D-Branes and Discrete Torsion
II", hep-th/9903031.}. 
Note that this map allows one to determine the effective BPS 
charges of the Gepner model boundary states as seen by a low energy 
observer. For B type boundary states 
\ref\OOY{H. Ooguri, Y. Oz, Z. Yin, ``D-Branes on Calabi-Yau Spaces and 
Their Mirrors'', Nucl. Phys. {\bf B477} (1996) 407, hep-th/9606112.}
(which correspond to 
D-branes wrapping even homology cycles), these are not the same as 
the microscopic 
D-brane charges, since the former include gravitational corrections
\nref\D{M.R. Douglas, ``Branes within Branes'', contributed to
``Cargese 1997, Strings, Branes and Dualities'', 267-275,
hep-th/9512077.}%
\nref\GHM{M. Green, J.A. Harvey and G. Moore,
``I-Brane Inflow and Anomalous Couplings on D-Branes'',
Class. Quant. Grav. {\bf 14} (1997) 47, hep-th/9605033.}%
\nref\HM{J.A. Harvey and G. Moore, ``On The Algebras of BPS States'',
Commun. Math. Phys. {\bf 197} (1998) 489, hep-th/9609017.}%
\nref\CY{Y.-K E.Cheung and Z. Yin, ``Anomalies, Branes and Currents'',
Nucl. Phys. {\bf B517} (1998) 69, hep-th/9710206.}%
\refs{\D,\GHM,\HM,\MM,\CY}. In such cases, the relation between 
the microscopic and effective charges involves the computation of a 
Mukai vector. The question of relating physical BPS states 
in the two regimes is more subtle and it requires a more detailed 
knowledge of the geometry of the Calabi-Yau variety near the 
large radius limit, including detailed results on the 
classification of vector bundles and special lagrangian cycles. 

In order to obtain concrete results, we focus on B type boundary states 
in the two parameter model ${\bf P}_4^{1,1,1,6,9}[18]$  of 
\ref\candb{P. Candelas, A. Font, S. Katz and D.R. Morrison,
``Mirror symmetry for two parameter models. 2,''
Nucl. Phys. {\bf B429}, 626 (1994) hep-th/9403187.}.
After resolving the singularities, these hypersurfaces exhibit a 
structure of elliptically fibered Calabi-Yau varieties which 
facilitates the construction of the relevant holomorphic bundles 
(or sheaves). 
Exploiting this feature, we explicitly analyze moduli and  
stability questions of Gepner model BPS states in the large radius limit. 
Similar issues have been addressed in \refs{\bdlr}
for the quintic and 
\ref\DG{D.-E. Diaconescu and J. Gomis, ``Fractional Branes and Boundary
States in Orbifold Theories'', hep-th/9906242.}
for local orbifold models. 

At the Gepner point, the boundary states are generically organized in
orbits of a discrete symmetry group (in the present model 
${\bf Z}_{18}$, as detailed in section four). 
We find that, within a given orbit, certain Gepner model states 
become unstable in the geometric region, signaling the crossover of 
a wall of marginal stability. At the same time, an important 
fraction of states in the same orbit are supersymmetric and stable 
in the large radius limit. This shows that the discrete ${\bf Z}_{18}$ 
symmetry is not a good symmetry in the geometric phase. 

Perhaps the most striking examples consist 
of the nonsupersymmetric D6 + D0 and respectively 
D4 + D2 states, 
which are repulsive in the large volume limit. On the other hand, 
they can be shown to correspond to supersymmetric Gepner model 
boundary states, giving explicit examples for the transitions 
predicted in \bdlr. This result is 
particularly interesting when interpreted in terms of D3-branes 
wrapping middle homology cycles in the mirror manifold. 
According to 
\ref\SYZ{A. Strominger, S.-T. Yau, E. Zaslow, ``Mirror Symmetry is 
T-Duality'', Nucl. Phys. {\bf B479} (1996) 243, hep-th/9606040.}, 
mirror symmetry can be thought as 
T-duality on the $T^3$ fibers of a special lagrangian fibration. 
A Calabi-Yau hypersurface $X$ near the 
large radius limit is mapped 
to a mirror manifold ${\hat X}$ in a neighborhood of the 
large complex structure limit. 
This transformation maps the D0-brane to a D3-brane wrapping the 
${\hat T}^3$ fibers of the dual fibration, while the  D6-brane is mapped 
to a D3-brane wrapping the base $B$ of the fibration. 
Therefore our analysis shows that for ${\hat X}$ in a certain 
neighborhood of the large complex structure limit, there should 
not exist any special lagrangian cycle in the homology class 
${\hat T}^3 + B$. However, in a different region of the complex 
structure moduli space of ${\hat X}$, 
that maps to a neighborhood of the Gepner point, the same 
homology class should actually contain a special lagrangian 
cycle that corresponds to the supersymmetric Gepner model boundary 
state. This provides a concrete example, in a compact Calabi-Yau 
space, for the transitions discussed in 
\ref\DJ{D. Joyce, ``On counting special Lagrangian homology 3-spheres'', 
hep-th/9907013.}
and from a physical point of view in 
\ref\KM{S. Kachru and J. McGreevy, 
``Supersymmetric Three-cycles and (Super)symmetry Breaking'',
hep-th/9908135.}.
It would be very interesting to understand the 
mathematical details of this transition in the present context, but we 
leave this for future work. 

For the boundary states that correspond to supersymmetric brane 
configurations, we find a remarkable agreement between the number 
of moduli computed in the two regions. This suggests that the 
superpotential couplings considered in \bdlr\ vanish in this model, 
but we do not check this explicitly. Also, an interesting point is 
the presence of a Gepner model boundary state with the charge of a 
single D0-brane (unlike the quintic studied in \bdlr).\ This shows that 
D0-branes are not necessarily a sign of geometrical compactification. 

The paper is structured as follows. Section two consists of a 
brief review of classical and quantum aspects of 
the two parameter model of \candb.\ In section three, we explain the 
relation between microscopic D-branes and BPS states on the moduli
space. Section four is devoted to boundary states in the Gepner 
model, emphasizing the SCFT intersection form and marginal 
deformations. In section five we tie together all loose ends 
and construct an explicit map between Gepner boundary states 
and D-branes. 

\newsec{The Geometry of the Elliptic Model}

This section consists of a brief review of the classical and quantum 
geometry of the two parameter model of \candb.\ We focus on facts of 
direct relevance to the spectrum of BPS states in a neighborhood of 
the Gepner point and respectively the large radius limit. 

\subsec{Classical Geometry}

The elliptic model $\BP^{(1,1,1,6,9)}_4[18]$ describes degree $18$ 
hypersurfaces in the weighted projective space $\BP^{(1,1,1,6,9)}_4$. 
A simple example of such a hypersurface $X$ is given by the equation
\eqn\surf{
z_1^{18}+z_2^{18}+z_3^{18}+z_4^3+z_5^2=0}
in the homogeneous coordinates
\eqn\homog{
(z_1,z_2,z_3,z_4,z_5)=
(\lambda z_1,\lambda z_2,\lambda z_3,\lambda^6 z_4,\lambda^9 z_5).}
>From this representation, it is easy to see that 
$\BP^{(1,1,1,6,9)}_4$ has a singular line along $z_1=z_2=z_3=0$, 
which intersects $X$ in a single point. Blowing up this singular 
line gives an exceptional divisor $E$ in $X$. Another divisor $S$ 
is defined by the first order polynomials in $x_1$, $x_2$ and $x_3$. 
These two divisors generate $H_4(X,\BZ)$. 
The elliptic fibration structure is induced by the linear 
system $|S|$ generated by $z_1$, $z_2$ and $z_3$ which 
maps $X$ to $\BP^2$. This fibration has a section given by the 
exceptional divisor $E$. The generic fiber can be proved to be an
elliptic curve, whose homology class in $H_2(X)$ will be denoted by $h$. 
The second generator $l$ of $H_2(X,\BZ)$ is the hyperplane class of $E$. 
For further use, we record the intersection relations
\eqn\inters{\eqalign{
& h= S^2,\qquad l=E\cdot S,\cr
& h\cdot E=1,\qquad h\cdot S=0,\qquad l\cdot E=-3,
\qquad l,\cdot S=1\cr
& E^2=-9,\qquad E^2\cdot S=-3,\qquad E\cdot S^2=1,\qquad S^3=0.\cr}}
We choose the generators\foot{Note that this choice is different from 
that of \candb\ where the generators are $H=E+3S$ and $S$. Therefore,
their coordinates are related by a linear transformation to ours. 
The present choice is motivated by the relation to D-brane states 
which will be explained latter.} of the K\"ahler cone to be $(E,S)$, 
so that a generic K\"ahler class is written $J=t_1E+t_2S$,
where $(t_1,t_2)$ are classical coordinates on the K\"ahler moduli space.

\subsec{Quantum Geometry}

Since the classical structure of the K\"ahler moduli space of $X$
is deformed by stringy $\alpha^\prime$ corrections, exact results 
can be obtained by considering the complex structure moduli space 
of the mirror ${\hat X}$. As explained in \candb,\ the mirror family 
can be obtained by applying the Greene-Plesser construction
\nref\grpl{B.R. Greene and M.R. Plesser,``Duality In Calabi-Yau Moduli
Space,'' Nucl. Phys. {\bf B338} (1990) 15.}%
\refs{\grpl}.
This results in a two parameter family of complex varieties of the 
form 
\eqn\msurf{
z_1^{18}+z_2^{18}+z_3^{18}+z_4^3+z_5^2-
18\psi z_1z_2z_3z_4z_5-3\phi x_1^6x_2^6x_3^6 =0.}
The complex parameters $(\phi,\psi)$ are subject to certain discrete 
identifications, therefore they are actually coordinates on a cover 
of the moduli space.

The exact special geometry of this model is described by a six-vector 
of periods 
$\Pi\equiv(\Pi_1\ldots \Pi_5)^{t}=
\left({\cF^0, \cF^1, \cF^2,1, t_1, t_2}\right)^{t}$
where $\cF$ is the $N=2$ prepotential. Using complex 
inhomogeneous coordinates 
$(t_1, t_2)$ in a neighborhood of the large radius limit, we have 
the following asymptotic expansion \candb\
\eqn\asympexpA{
\left[\matrix{\Pi_1\cr \Pi_2\cr \Pi_3\cr \Pi_4\cr \Pi_5\cr \Pi_6\cr}
\right]=
\left[\matrix{\cF^0\cr \cF^1\cr \cF^2\cr 1\cr t_1\cr t_2\cr}\right]\sim
\left[\matrix{{1\over 2}\left(3t_1^3+3t_1^2t_2+t_1t_2^2\right)
+{17\over 4}t_1+{3\over 2}t_2\cr
-{1\over 2}t_2^2+{3\over 2}t_2-{1\over 4}\cr
-{1\over 2}\left(3t_1^2+2t_1t_2\right)+{3\over 2}t_1+{3\over 2}\cr
1\cr t_1\cr t_2\cr }\right].}
Note that $\Pi$ represents the vector of periods of the holomorphic
three-form ${\hat \Omega}$ on the mirror manifold ${\hat X}$ 
with respect to an integral basis of three-cycles. Due to the different 
choice of generators of the K\"ahler cone, this basis is not canonical 
symplectic as in \candb.\ In particular, the intersection form is 
given by 
\eqn\lgint{
I_L=\left(\matrix{&0 &0 &0 &-1 &0 &0 \cr
                  &0 &0 &0 &0  &-1 &-3 \cr
                  &0 &0 &0 &0  &0 &-1\cr
                  &1 &0 &0 &0  &0 &0\cr
                  &0 &1 &0 &0  &0 &0\cr
                  &0 &3 &1 &0  &0 &0\cr}\right).}
In order to interpolate between Gepner model boundary states and 
large radius limit branes, the periods \asympexpA\ must be analytically 
continued to a neighborhood of the Gepner point. The natural basis of 
periods in this region is described by an overcomplete eighteen-vector 
$(\omega_0,\ldots,\omega_{17})^{t}$ whose entries are cyclically permuted 
by the quantum ${\bf Z}_{18}$ discrete symmetry of the Gepner model. 
These periods satisfy the relations 
\eqn\perrel{\eqalign{
&\omega_i+\omega_{i+9}=0,\cr
&\omega_i-\omega_{i+3}+\omega_{i+6}=0,}}
which leave only six independent periods $(\omega_0,\cdots,\omega_5)^{t}$.
The periods $\Pi$ and $\omega$ are related by analytic continuation, 
resulting in $\Pi=m\omega$ with the connection matrix
\eqn\lgtrans{
m=\left(\matrix{ -1 & 1 & 0 & 0 & 0 & 0 \cr
                 1 & 0 & 0 & -1 & 1 & 0 \cr
                 0 & 1 & 1 & 1 & 0 & 0 \cr
                 1 & 0 & 0 & 0 & 0 & 0 \cr
                 -1 & 0 & 0 & 1 & 0 & 0 \cr
                 2 & 0 & 0 & -2 & 1 & 1 \cr}\right).}
With respect to the basis of periods $(\omega_0,\ldots,\omega_5)^{t}$, 
the intersection form on $H^3({\hat X},\bf Z)$ takes the form 
\eqn\ginter{
I_G=m^{-1}I_L\,m^{-1\,t}=\left(
\matrix{ 0 & 1 & -1 & 0 & 0 & 0 \cr
         -1 & 0 & 1 & -1 & 0 & 0 \cr
         1 & -1 & 0 & 1 & -1 & 0 \cr
         0 & 1 & -1 & 0 & 1 & -1 \cr
         0 & 0 & 1 & -1 & 0 & 1 \cr
         0 & 0 & 0 & 1 & -1 & 0 \cr}\right).}
Using the relations \perrel,\ the matrix $I_G$ can be expressed in terms 
of a ${\bf Z}_{18}$ shift matrix 
\eqn\sinter{
I_G=-g^{17}(1-g^{17})(1-g^{12})(1-g^9).}
This particular form of the intersection matrix will play an important 
role in the interpolation between the Gepner point 
and the large radius limit. 

Finally, let us note that the large radius point sits at the intersection
of two divisors $G_1$, $D_{\infty}$ on the boundary of the moduli space. 
\candb.\ The monodromy matrices about the two divisors, expressed with 
respect to the basis of periods \asympexpA,\ read
\eqn\lrmon{
S_1=\left(\matrix{& 1 & -1 & -3 & 10 & 9 & 3\cr
                  & 0 & 1 & 0 & 0 & 0 & 0 \cr
                  & 0 & 0 & 1 & 0 & -3 & -1\cr
                  & 0 & 0 & 0 & 1 & 0 & 0\cr
                  & 0 & 0 & 0 & 1 & 1 & 0\cr
                  &0 & 0 & 0 & 0 & 0 & 1\cr}\right)
\quad 
S_2=\left(\matrix{& 1 & 0 & -1 & 3 & 2 & 0\cr
                  & 0 & 1 & 0 & 1 & 0 & -1\cr
                  & 0 & 0 & 1 & 0 & -1 & 0\cr
                  & 0 & 0 & 0 & 1 & 0 & 0\cr
                  & 0 & 0 & 0 & 0 & 1 & 0\cr
                  & 0 & 0 & 0 & 1 & 0 & 1\cr}\right).}
These monodromy transformations correspond to shifting the B-field 
by the integral cohomology classes $H=E+3S$ and respectively $S$. 
In the next section they will be reinterpreted as natural automorphisms 
of the K theory group of $X$. This is a simple particular 
case of a more ambitious 
program which proposes an identification between the derived category 
of complexes of sheaves of $X$ and the Fukaya-Floer category of special 
lagrangian submanifolds of ${\hat X}$. 
\nref\K{M. Kontsevich, ``Homological Algebra of Mirror Symmetry'', 
alg-geom/9411018.}%
\nref\PZ{A. Polishchuk and E. Zaslow, 
``Categorical Mirror Symmetry: The Elliptic Curve'', alg-geom/9801119.}%
\nref\CVb{C. Vafa, ``Extending Mirror Conjecture to Calabi-Yau with 
Bundles'', hep-th/9804131.}%
\nref\BSG{C. Bartocci, U. Bruzzo and G. Sanguinetti, ``Categorial
mirror symmetry for K3 surfaces'', math-ph/9811004.}%
\refs{\K,\PZ,\CVb,\BSG}.
A systematic and general approach 
will appear in 
\ref\PH{Paul Horja, dissertation thesis, Duke University.}.

\newsec{D-branes and Periods}

In this section we explore the relation between the spectrum of 
BPS states of this model and microscopic D-brane states in the 
large volume limit. Before discussing the technical aspects of the 
problem, a couple of guiding remarks are in order. The BPS charge 
lattice of the low energy effective theory is an integral symplectic 
lattice which can be identified with the middle cohomology lattice of 
the mirror manifold $H^3({\hat X},{\bf Z})$. 
The central charge corresponding to an integral vector 
$(n_6, n_4^1, n_4^2, n_0, n_2^1, n_2^2)$ is 
\eqn\zchargeA{
Z=n_6\Pi_1+n_4^1\Pi_2+n_4^2\Pi_3+n_0\Pi_4+n_2^1\Pi_5+n_2^2\Pi_6.}
On the other hand, in the large radius 
limit, the lattice of microscopic D-brane charges 
is an integral quadratic lattice identified with the K theory lattice 
$K(X)$. The map between these lattices is a nontrivial question 
in mirror symmetry, being related to the current efforts of extending 
the mirror principle to Calabi-Yau spaces with bundles. 
In the present case, we will construct a map between the 
low energy charges ${\bf n}$ and the topological invariants of the 
K theory class $\eta$ by exploiting the 
exact form of D-brane Chern-Simons couplings worked out in 
\refs{\D,\GHM,\HM,\MM,\CY}. The topological invariants of $\eta$ are 
simply given by the Chern character $\hbox{ch}(\eta)$.
The effective charges of a D-brane state represented by
$\eta$ are measured by the Mukai vector 
$Q\in H^0(X)\oplus H^2(X)\oplus H^4(X)\oplus H^6(X)$ given by 
\eqn\mukai{
Q=\cn{\eta}\sqrt{\td{X}}.}
The central charge associated to this state is then\foot{Note that 
our conventions are such that D0-brane configurations have 
$\int_X\hbox{ch}_3<0$. The associated K theory class is $-[{\co}_P]$ 
rather than $[\co_P]$, where $\co_P$ is a skyscraper sheaf of length
one supported 
at the point $P$. Similarly, a D2-brane wrapping a curve $C\subset X$ 
is represented by the K theory class $-[\co_C]$ rather than $[\co_C]$.}
\eqn\zchargeB{
Z(t)={t^3\over 6}Q^0-{t^2\over 2}Q^2+tQ^4-Q^6.}
The comparison of \zchargeA\ and \zchargeB\ gives the relation 
between the low energy charges and the topological invariants of $\eta$.
We derive next explicit formulae for the cases when $\eta$ describes 
either D6-branes wrapped on $X$ or D4-branes wrapped on submanifolds
of $X$. More general situations (for example if $\eta$ is not
representable by a sheaf on $X$) can be treated similarly. 

\subsec{The D6-Brane}

We now consider D-brane systems with nonzero D6-charge which can be 
represented by holomorphic vector bundles on $X$. In fact, it will turn 
out that this condition is too restrictive and we will actually have 
to enlarge the class of geometrical objects to coherent sheaves on $X$. 
Note also that the corresponding D-brane configuration is supersymmetric
only if $V$ is a stable sheaf\foot{We will not give the explicit 
definition of stability here. The Donaldson-Uhlenbeck-Yau theorem shows 
that stability is essentially equivalent with self-duality of the gauge 
field configuration, which is more familiar to physicists.}
\refs{\HM}.
Expanding \mukai,\ we obtain 
\eqn\chexpC{
Q=\left(r,c_1(V),\ch{2}{V}+{r\over 24}c_2(X),\ch{3}{V}
+{1\over 24}c_1(V)c_2(X)\right).}
The associated central charge formula is therefore 
\eqn\chargeC{
Z(Q)={r\over 6}t^3-{1\over 2}\ch{1}{V}t^2+
\left(\ch{2}{V}+{r\over 24}c_2(X)\right)t-\left(\ch{3}{V}+
{1\over 24}c_1(V)c_2(X)\right).}
A direct comparison of \zchargeA\ and \chargeC\ yields 
\eqn\charC{\eqalign{
& r=n_6\cr
& \ch{1}{V}=n_4^1E+n_4^2S\cr
& \ch{2}{V}=\left({3\over 2}n_4^2+n_2^1\right)h+
\left({3\over 2}n_4^1+n_2^2\right)l\cr
& \ch{3}{V}= -n_0
+{1\over 2}n_4^1-3n_4^2.\cr}}
Therefore
\eqn\charD{\eqalign{
& r=n_6\cr
& c_1(V)=n_4^1E+n_4^2S\cr
&c_2(V)=\left[{1\over 2}(n_4^2)^2-{3\over 2}n_4^2-n_2^1\right]h+\cr
&\qquad\quad\ \ \left[-{3\over 2}(n_4^1)^2+n_4^1n_4^2-
{3\over 2}n_4^1-n_2^2\right]l\cr
& c_3(V)={1\over 2}\left(3(n_4^1)^3-3(n_4^1)^2n_4^2+n_4^1(n_4^2)^2\right)
+{9\over 2}(n_4^1)^2-3n_4^1n_4^2-\cr
&\qquad\qquad n_4^1n_2^1-(n_4^2-3n_4^1)n_2^2-6n_4^2+n_4^1-2n_0.\cr}}
These formulae relate the topological invariants of the sheaf $V$ 
to the effective charges ${\bf n}$, as promised. In the next sections,
they will play an important role in identifying the geometric 
properties of the Gepner model boundary states. 

Before moving on to more general D-brane configurations, note that 
we can use the above explicit formulae to convert the monodromy 
transformations \lrmon\ into automorphisms of $K(X)$, as promised 
before. More precisely, we claim that the two monodromy transformations
correspond to the following automorphisms 
\eqn\Kautom{
[V]\rightarrow \left[V\otimes \co_X(H)\right],\qquad
[V]\rightarrow \left[V\otimes \co_X(S)\right],}
where $V$ is a coherent sheaf on $X$. 
To show this, note that tensoring by a line bundle $\co_X(D)$ 
changes the topological invariants of $V$ as follows 
\eqn\chshiftA{\eqalign{
& r\left(V^\prime\right)=r(V)\cr
& \ch{1}{V^\prime}=\ch{1}{V}+rD\cr
& \ch{2}{V^\prime}=\ch{2}{V}+\ch{1}{V}D+{r\over 2}D^2\cr
& \ch{3}{V^\prime}=\ch{3}{V}+\ch{2}{V}D+{1\over 2}\ch{1}{V}D^2+
{r\over 6}D^3\cr}}
where $V^\prime\simeq V\otimes \co_X(D)$.
Using \charC,\ a direct computation shows that the linear transformations 
acting on ${\bf n}$ corresponding to $D=H,S$ are precisely given by 
\eqn\lintransf{
M(H)={S_1}^{-1},\qquad M(S)={S_2}^{-1},}
where $S_1$, $S_2$ are the monodromy matrices in \lrmon. 
This proves the claim. 

\subsec{D4-Brane}

A different class of D-brane states can be obtained by wrapping lower 
even dimensional branes on submanifolds of $i:D\hookrightarrow X$. 
In order to obtain supersymmetric configurations, $D$ has to be a 
holomorphic cycle
\nref\BBS{K. Becker, M. Becker and A. Strominger, 
``Fivebranes, Membranes and Non-Perturbative String Theory'',
Nucl. Phys. {\bf B456} (1995) 130, hep-th/9507158.}
\refs{\BBS,\OOY}.
Multiple brane configurations are described 
as above by a coherent sheaf $V$ on $D$, which is required to 
be stable. The associated K theory class in $K(X)$ 
is simply defined by the torsion coherent sheaf $i_{\star}V$ 
which is the extension of $V$ by zero to $X$. Then, the Mukai vector 
can be computed by a simple application of the Grothendieck-Riemann-Roch 
formula for the embedding $i:D\hookrightarrow X$
\eqn\grrA{
i_*\left(\cn{V}\td{D}\right)=\cn{i_*V}\td{X}.}
Note that this formula relates the topological invariants of the sheaf 
$V$ on $D$ to the topological invariants of the torsion sheaf 
$i_{\star}V$ on $X$. 
A direct expansion yields 
\eqn\chexpC{\eqalign{
& \ch{1}{i_*V}= rD\cr
& \ch{2}{i_*V}= i_*c_1(V)+{r\over 2}i_*c_1(D)\cr
& \ch{3}{i_*V}= i_*\left(\ch{2}{V}+{1\over 2}c_1(V)c_1(D)+{r\over 12}
(c_1(D)^2+c_2(D))\right)-{r\over 12}Dc_2(X).\cr}}
It follows that the Mukai charge vector is 
\eqn\mukaiB{\eqalign{
Q=\biggl(0,rD,i_*c_1(V)+{r\over 2}i_*c_1(D),
&\ch{2}{V}+{1\over 2}c_1(V)c_1(D)+\cr
&{r\over 12}(c_1(D)^2+c_2(D))
-{r\over 24}Dc_2(X)\biggr).\cr}}
In the last entry of the above formula, the top Chern classes are 
evaluated on the corresponding fundamental cycles, resulting in numerical 
invariants. Using an adjunction formula
\eqn\form{Dc_2(X)=c_2(D)-D^3=c_2(D)-c_1(D)^2,}
the charge vector can be rewritten 
\eqn\mukaiC{
Q=\left(0,rD,i_*c_1(V)+{r\over 2}i_*c_1(D),
\ch{2}{V}+{1\over 2}c_1(V)c_1(D)+{r\over 8}c_1(D)^2+{r\over 24}
c_2(D)\right).}
Therefore, in the large volume limit, the central charge 
associated to the charge vector \mukaiB\ reads
\eqn\chargeA{\eqalign{
Z(Q)=&-{r\over 2}t^2D+\left(i_*c_1(V)+{r\over 2}i_*c_1(D)\right)t
-\ch{2}{V}-\cr
&{1\over 2}c_1(V)c_1(D)-{r\over 8}c_1(D)^2-{r\over 24}c_2(D).\cr}}
Note that $i_*c_1(V),\ i_*c_1(D)$ can be thought as curve classes 
on $X$ by Poincar\'e duality, therefore the intersection number 
in the second term is well defined.
With an eye on applications, we derive explicit formulae for multiple 
D4-branes with lower induced charges wrapping the cycles 
$E$ and $S$ on the elliptic fibration.

Branes wrapped on $E$ correspond to BPS states with charge vectors 
of the form\foot{Note that the allowed lower D2-brane charges 
correspond to cycles that can be embedded in $E$.}
${\bf n}=\left(0,n_4^1,0,n_0,0,n_2^2\right)$ with central 
charge
\eqn\chargeE{
Z({\bf n})=n_4^1\cF^1+n_2^2t_2+n_0.}
The topological invariants of the sheaf $V$ can be obtained from 
equations \chargeA\ and \chargeE\
\eqn\charE{\eqalign{
&r=n_4^1\cr
&\ch{1}{V}=n_2^2l\cr
&\ch{2}{V}=-n_4^1-{3\over 2}n_2^2-n_0.\cr}}
Therefore
\eqn\charF{\eqalign{
&r=n_4^1\cr
&c_1(V)=n_2^2l\cr
&c_2(V)={1\over 2}n_2^2(n_2^2+3)+n_4^1+n_0.\cr}}
Similarly, branes wrapping $S$ correspond to BPS charge vectors of the 
form  
${\bf n}=\left(0,0,n_4^2,n_0,n_2^1,n_2^2\right)$ with central charge 
\eqn\chargeF{
Z({\bf n})=n_4^2\cF^2+n_2^1t_1+n_2^2t_2+n_0.}
Using again \chargeA\ and \chargeF,\ we deduce
\eqn\charG{\eqalign{
& r=n_4^2\cr
& \ch{1}{V}=(2n_4^2
+n_2^1)h+n_2^2l\cr
& \ch{2}{V}=-3n_4^2+{1\over 2}n_2^2-n_0.\cr}}
Therefore
\eqn\charH{\eqalign{
&  r=n_4^2\cr
& c_1(V)=(2n_4^2+n_2^1)h+n_2^2l\cr
& c_2(V)=-{1\over 2}n_2^2(3n_2^2+1)+2n_2^2n_4^2+
n_2^1n_2^2+3n_4^2+n_0.\cr}}
More general configurations involving various configurations of 
D4-branes and D2-branes on holomorphic cycles in $X$ can be treated 
similarly. 

\newsec{Boundary States in the Gepner Model}

As explained before, the Gepner point is a special point in the moduli
space where the superconformal field theory is exactly 
solvable \refs{\wiA,\lvw,\wiB}. This feature allows us to obtain 
valuable information on the spectrum of BPS states in a deep stringy 
regime of the theory. 
In the present section, we explain the construction of B type boundary 
state in the $(k=16)^3(k=1)$ Gepner model which is continuously connected 
to the elliptic Calabi-Yau compactification considered in the previous 
sections. The construction is closely related to the one applied 
to rational conformal field theories in \cardy\ and it has been 
considered in detail in \refs{\reso,\GSa,\GSb}. Our approach is focused 
on the conformal field theory computation of the symplectic intersection
form on the BPS charge lattice \dofi.\ As in \bdlr,\ this is a crucial 
ingredient in the geometric interpretation of Gepner model boundary
states. 

The B type boundary states at the Gepner point $(k=16)^3(k=1)$
are labeled as in \bdlr\ by $|L_1,L_2,L_3,L_4; M;S\rangle$ 
where $0\leq L_1,L_2,L_3\leq 8$, $L_4=0$, $M\in \BZ_{36}$
\eqn\bdst{
M=\sum_{i=1}^4{K^\prime M_i\over k_i+2}}
with $K^\prime = \hbox{l.c.m.}\{k_i+2\}$ and $S\in 2\BZ_2$.
For fixed $L=(L_i)$, the states with different $(M,S)$ form an 
orbit under the $\BZ_{18}$ discrete symmetry group and the two values of 
$S$ correspond to a brane and the corresponding antibrane. 

\subsec{The Intersection Form}

A first step in deriving a geometric interpretation of the Gepner model 
boundary states is to calculate their intersection numbers using exact 
conformal field theory techniques. As explained in 
\nref\BD{M. Berkooz and M. Douglas, ``Five-branes in M(atrix) theory,''
Phys. Lett. {\bf B395}, (1997) 196, hep-th/9610236.}%
\refs{\BD,\dofi,\bdlr}, 
these numbers can be obtained by computing $\hbox{tr}(-1)^{F}$ in the 
internal part of the open string R sector. Since this quantity is an 
index, it is unchanged under marginal deformations of the SCFT on the 
moduli space. Therefore it can be reliably compared with the large 
radius limit intersection matrix given by \ginter\ in a basis adapted 
to the ${\bf Z}_{18}$ symmetry of the Gepner point. An explicit 
calculation gives
\eqn\btrace{
	I_B={1\over C}(-1)^{{{S-\tilde S}\over 2}}\sum_{m_j'}
	\delta^{(K')}_{{{M-\tilde M}\over 2}+
		\sum{K'\over{2k_j+4}}(m_j'+1)}
	\prod_{j=1}^r N^{m_j'-1}_{L_j,\tilde L_j}\ ,}
where $N^l_{L,\tilde L}$ are the $SU(2)_k$ fusion rule coefficients. 
This formula is very cumbersome, but it can be rewritten in a much 
simpler form. To this end one can note that the states within one 
$\BZ_{18}$ orbit ($L_j$ are fixed) can be labeled by a 36 dimensional 
row vector $q_B$ with all entries equal to zero, except for the 
$M$th entry, which equals one.

In this notation the intersection matrices for boundary states in 
fixed orbits of the $\BZ_{18}$ symmetry can be expressed in terms of 
shift matrices $g$. Each factor $N^{m_j'-1}_{L_j,\tilde L_j}$ 
in \btrace\ can be replaced in matrix notation by a factor 
\foot{The matrix $g^\half$ is understood as the basic 36-dimensional 
shift matrix.}
\eqn\fusA{
n_{L,\tilde L}=n_{\tilde L,L}=
g^{{|L-\tilde L|}\over 2}+g^{{{|L-\tilde L|}\over
2}+1}+\cdots+g^{{L+\tilde L}\over 2}
-g^{-1-{{|L-\tilde L|}\over 2}}-\cdots-g^{-1-{{L+\tilde L}\over 2}}.
}
The delta function constraint is a shifted $U(1)$ projection, showing 
that the intersection matrix is $\BZ_{18}$ invariant.

As specified before, the BPS charge vectors associated to the Gepner 
model boundary states can be found by comparing the intersection matrices
\btrace\ and \sinter.\ It turns out that it is more convenient to 
work in a different basis for boundary states, related to the present one 
by the linear transformation ${1\over \sqrt{2}}\left(1-g^9\right)$.
In the new basis, the matrix \btrace\ reads
\eqn\binter{
I_B=(1-g^{17})^3(1-g^{12})(1-g^9).}
This is antisymmetric and easy combinatorics shows that 
it's rank is $6$. From \sinter\ and \binter\ it can be seen that the two 
matrices $I_G$ and $I_B$ are related by
\eqn\gbtrans{
I_B=(1-g)I_G(1-g)^t.}
This connects the series of boundary states with $(L_j)=(0,\cdots,0)$ 
to the basis of periods $\omega$ at the Gepner point.

To find the charges of boundary states with $L_j\ne 0$ one has to find 
a linear transformation $t_L$ which generates the different factors 
$n_{L,\tilde L}$ from $n_{0,0}=(1-g^{-1})$. The change of basis is 
expressible in terms of $g$
\eqn\deft{
t_L=t_L^t=\sum_{l=-{L\over2}}^{L\over 2}g^l.}
The first step in verifying this is to relate $n_{L,0}$ to $n_{0,0}$
\eqn\nrelA{
t_Ln_{0,0}=\sum_{l=-{L\over 2}}^{L\over 2}g^l(1-g^{-1})=
\sum_{l=-{L\over 2}}^{L\over 2}g^l-
\sum_{l=-{L\over 2}-1}^{{L\over 2}-1}g^l=
g^{L\over 2}-g^{-{L\over 2}-1}=n_{L,0}.}
Multiplying this with $t_{\tilde L}$ one gets
\eqn\nrelB{
t_Ln_{0,0}t_{\tilde L}^t=\sum_{l=-{L\over 2}}^{L\over 2}g^l
(g^{\tilde L\over 2}-g^{-{\tilde L\over 2}-1})=
\sum_{l={|L-\tilde L|\over 2}}^{L+\tilde L\over 2}(g^l-g^{-l-1})=
n_{L,\tilde L}.}

The charge of the boundary state $q_B$ in the Gepner basis is then 
given by $q_G=q_Bt_{L_1}t_{L_2}t_{L_3}(1-g)$. This row vector has 
only entries in the even columns, which means that all the odd columns 
can be omitted, leaving a $18$ dimensional vector. There are only $6$ 
independent charges and the relations \perrel\ can be used to reduce 
$q_G$ to it's first $6$ entries. The large radius charge $q_L$ is 
then easily calculated from the reduced $q_B$ by $q_L=q_Gm$. 

\subsec{Marginal Operators}

For comparison with geometric results it will be interesting to compute
the number of boundary marginal operators. As explained in \bdlr\ they 
can be expressed in terms of the matrices 
$\tilde n_{L,\tilde L}=|n_{L,\tilde L}|$. The number of boundary 
marginal operators for only one boundary state $\ket{L_j,M,S}$ 
is given by the diagonal part of
\eqn\nmarg{
\half\tilde n_{L_1,L_1}\tilde n_{L_2,L_2}
\tilde n_{L_3,L_3}(1+g^{12})(1+g^9)
-\#\hbox{vac}.}
The number of vacua is normally $1$ and for each $L_j=8$ it is  
multiplied by $2$. The following table shows the number of marginal 
operators for some important boundary states
\eqn\margdef{
\matrix{(L_1,L_2,L_3)&\#(\hbox{marg})&\#(\hbox{vac})\cr
	(0,0,0)&0&1\cr
	(1,0,0)&2&1\cr
	(2,0,0)&3&1\cr
	(3,0,0)&3&1\cr
	(4,0,0)&3&1\cr
	(5,0,0)&3&1\cr
	(6,0,0)&4&1\cr
	(7,0,0)&6&1\cr
	(8,0,0)&6&2}}
These boundary operators are massless, but they might have a 
superpotential, with flat directions corresponding to the truly 
marginal operators.

If the number of vacua in the open string sector is different from zero, 
e.g. two, one might think of the boundary state as two different 
D-branes. This would fit with the picture of a Coulomb branch in the 
world volume theory in which the gauge group is $U(1)^2$. The reason 
that these boundary states appear in the formalism of the Gepner model 
as a single boundary state could be related to 
the higher symmetry algebra that 
these boundary states respect.

\newsec{Boundary States and Vector Bundles}

In this final section, we collect the results obtained so far and 
establish an explicit connection between Gepner model D-branes 
and supersymmetric brane configurations on the elliptically 
fibered Calabi-Yau variety. To summarize, this process essentially 
involves two stages. First, we interpret the boundary states at the 
Gepner point as generic BPS states on the moduli space. The next 
step involves a translation of the BPS charge vectors ${\bf n}$ into 
microscopic D-brane charges using the Chern-Simons couplings as 
explained in section three. Note that this procedure involves analytic 
continuation between two distinct regions of the moduli space, 
therefore the spectrum of BPS states may be affected by jumping 
and marginal stability phenomena. We will not attempt to give
a comprehensive 
study of these issues here, but the geometric picture will 
eventually provide significant information on the stability of the 
Gepner model D-branes in the large radius limit. Moreover, we will 
also compare the number of moduli of a given D-brane configuration 
in the two regimes, finding a remarkable agreement. This suggests that 
the marginal deformations found in section 4.2 are in fact truly
marginal and the corresponding flat directions are not lifted by 
superpotential couplings. Note that an exhaustive treatment of all
Gepner boundary states is not possible due to their large number 
(1485, according to some combinatorial arguments). Therefore 
we will restrict in the following to a subset of states 
which admit a simple geometric interpretation. 

\subsec{$L=(0000)$}

The simplest series is $L=(0000)$. The 18 
states forming a ${\bf Z}_{18}$ orbit are grouped in D-brane/anti-D-brane 
pairs, therefore there are only 9 relevant charge vectors, which 
are listed in the following table
\eqn\tableA{\matrix{&{\hbox{No}}
& n_6 & n_4^1 & n_4^2& n_0 & n_2^1 & n_2^2\cr
&{\bf n}_1& 1 & 0 & 0 & 0 & 0 & 0 \cr
&{\bf n}_2& 2 & 0 & -1 & 3 & 1 & 0 \cr
&{\bf n}_3& 1 & -1 & -1 & 3 & 2 & 1 \cr
&{\bf n}_4& 1 & -1 & 0 & 1 & 0 & 0 \cr
&{\bf n}_5& 1 & 0 & -1 & 3 & 2 & 0\cr
&{\bf n}_6& 0 & 1 & 0 & -1 & 0 & 0\cr
&{\bf n}_7& 0 & 2 & 0 & 0 & 0 & -1\cr
&{\bf n}_8& 0 & 1 & 0 & 0 & 0 &-1\cr
&{\bf n}_9& 2 & -2 & -1 & 3 & 1 & 1.\cr}}
The corresponding topological invariants can be computed by a direct 
application of the formulae in sections 3.1 and 3.2. We find that the
states ${\bf n}_1$, ${\bf n}_3$, ${\bf n}_4$, ${\bf n}_5$
correspond to the complex holomorphic line 
bundles $\co_X$, $\co_X(-E-S)$, $\co_X(-E)$, $\co_X(-S)$ respectively. 
These are clearly stable and describe supersymmetric single D6-brane 
configurations with induced anti-D4 charges. They are also rigid, 
since $h^{0,1}(X)=0$, therefore the number of moduli is zero, in 
agreement with the results in 3.5. Similarly, the states 
${\bf n}_6$ and ${\bf n}_8$ correspond to the holomorphic line bundles 
$\co_E$ and 
$\co_E(-l)$ on the section $E\simeq {\bf P}^2$ of the elliptic
fibration. These are again stable and rigid and correspond to single 
D4-branes on $E$. 

The remaining states, ${\bf n}_{2,7,9}$ are more interesting since 
they correspond 
to multiple branes. The first charge is associated to a vector bundle 
$V$ on $X$ with topological invariants 
\eqn\topA{
r(V)=2,\qquad c_1(V)=-S,\qquad c_2(V)=h,\qquad c_3(V)=0.}
A holomorphic vector bundle with these 
characteristics can be easily constructed as a pull back of a rank two 
vector bundle $W$ on the base ${\bf P}^2$, $V=\pi^*W$. For $W$, we find 
\eqn\basevb{
r(W)=2,\qquad c_1(W)=-l,\qquad c_2(W)=1}
and it turns out that this is an exceptional bundle\foot{Exceptional 
bundles on a surface are in general 
characterized by $H^1(\en(W))\simeq 0$ and $H^2(\en(W))\simeq 0$. 
On the projective plane, 
one can prove that this implies stability.} 
on the projective plane ${\bf P}^2$ 
\nref\DL{J.-M. Dr\'ezet and J. Le Poitier, ``Fibr\'es Stable et
fibr\'es Exceptionnels sur Le Plan Projectif'', Ann. Scient. Ec Norm. Sup.
{\bf 18} (1998) 105.}%
\nref\L{J. Le Poitier, ``Lectures on Vector Bundles'', Cambridge
University Press, 1997.}%
\refs{\DL,\L}.
In particular, $W$ is stable and rigid. 
We can prove that $V$ is also rigid as follows.
The infinitesimal deformations of $V$ on $X$ are parameterized by a
the cohomology group 
\eqn\defA{
H^1\left(X,\hbox{End}(V)\right)=H^1\left(X,V\otimes V^*\right)
=H^1\left(X,\pi^*(W\otimes W^*)\right).}
This can be evaluated using the Leray spectral sequence for 
$\pi:X\rightarrow B$
\eqn\spseq{
H^1\left(X,\pi^*(W\otimes W^*)\right)\simeq
H^0\left(B,W\otimes W^*\otimes K_B\right)\oplus 
H^1\left(B,W\otimes W^*\right).}
Using Kodaira-Serre duality, we have 
\eqn\vanA{\eqalign{
& H^0\left(B,W\otimes W^*\otimes K_B\right)
\simeq H^2\left(B,\en(W)\right)^*=0\cr
& H^1\left(B,W\otimes W^*\right)=H^1\left(B,\en(W)\right)=0,}}
since $W$ is exceptional. Therefore $V$ is indeed rigid and it can be 
proved similarly that $V$ is also simple i.e. it has no nontrivial 
automorphisms $\hbox{Hom}\left(V,V\right)\simeq {\bf C}$.
The stability of $V$ is harder to analyze and we have not been able to 
obtain a definite result. 

Next, it can be checked that the state ${\bf n}_7$ represents a D4-brane 
with multiplicity two wrapped on the section $E$. The associated vector 
bundle turns out to be in fact isomorphic to the exceptional bundle 
$W$ considered in the previous paragraph. This is rigid and stable, 
therefore we obtain a supersymmetric configuration with no moduli. 
Finally, the ninth state corresponds to a bundle $V$ with topological 
invariants 
\eqn\topB{
r(V)=2,\qquad c_1(V)=-2E-S,\qquad
c_2(V)=h-2l,\qquad c_3=0.}
A bundle with these topological invariants can be easily constructed 
as $V\simeq \pi^{\star}W\otimes \co_X(-E)$. It's properties are similar 
to those of $\pi^{\star}W$ i.e. $V$ is rigid and simple. 

To summarize the results, we have found that all states, except possibly 
${\bf n}_2$, ${\bf n}_7$ and ${\bf n}_9$, are stable and supersymmetric 
in the large radius limit. Moreover, the number of geometric moduli 
agrees with the number of marginal deformations of the boundary states. 
This provides supporting evidence for the decoupling of the bulk 
K\"ahler moduli argued
 in \bdlr.\ The states 
${\bf n}_2$, ${\bf n}_7$ and ${\bf n}_9$ are more intriguing since 
we have not been able to settle the issue of stability in the large 
radius limit. On the other hand, since the associated bundles are simple, 
these D-brane configurations should correspond to one particle states 
in the four dimensional effective theory. If the bundles turn
out to be unstable, it would be interesting to understand if these 
are stable non-BPS states in the large radius limit. 

Next, we consider another group of Gepner model boundary states 
which lead to more interesting physical configurations. 

\subsec{$L=(2000)$}

The charge vectors of the states in this series are 
\eqn\tableB{\matrix{&{\hbox{\bf n}}
& n_6 & n_4^1 & n_4^2& n_0 & n_2^1 & n_2^2\cr
&{\bf n}_1& 0 & 0 & 0 & 1 & 0 & 0 \cr
&{\bf n}_2& 0 & 0 & 0 & 0 & -1 & 0 \cr
&{\bf n}_3& 0 & 0 & 0 & 1 & -1 & 0 \cr
&{\bf n}_4& 1 & 0 & 0 & 1 & 0 & 0 \cr
&{\bf n}_5& 1 & -1 & 0 & 1 & -1 & 0\cr
&{\bf n}_6& 1 & 0 & -1 & 2 & 1 & 0\cr
&{\bf n}_7& 1 & -1 & -1 & 2 & 2 & 1\cr
&{\bf n}_8& 0 & 1 & 0 & 0 & 1 &0\cr
&{\bf n}_9& 0 & 1 & 0 & 0 & -1 & -1.\cr}}
The first charge vector in this series is particularly interesting since 
it corresponds to a D0-brane on $X$. Formally, this is described by a K 
theory class $\eta\in K(X)$ with topological invariants 
\eqn\topD{
r(\eta)=0,\qquad  \ch{1}{\eta}=0,\qquad
\ch{2}{\eta}=0,\qquad \ch{3}{\eta}=-1.}
This identifies $\eta$ as $-[\co_P]$ where $\co_P$ is a skyscraper sheaf 
of length one supported at the point $P\in X$. 
The next two states ${\bf n}_{2,3}$ also have a simple physical 
interpretation. They correspond to a D2-brane wrapped on the elliptic 
fiber and respectively to a D2-brane wrapped on the elliptic fiber 
with a magnetic flux inducing one unit of D0-charge. The K theory 
classes can be easily constructed. Pick $i:Y\hookrightarrow X$ to be an 
arbitrary smooth 
elliptic fiber in the class $h$ and pick $\cl_Y$ to be a holomorphic 
line bundle on $Y$. We must have $\hbox{deg}(\cl_Y)=0$ for ${\bf n}_2$ 
and $\hbox{deg}(\cl_Y)=1$ for ${\bf n}_3$. Then the required K theory 
class is $\eta=-[i_{\star}\cl_Y]$. 
All these configurations are supersymmetric and we can also determine 
the moduli space. For a D0-brane, the moduli space is simply isomorphic 
to $X$. For the D2-branes, the moduli are parameterized by a point 
in the base ${\bf P}^2$ representing the projection of the elliptic 
fiber $Y$ and the choice of a flat line bundle on $Y$ (specifying
the Wilson lines). Therefore the global moduli space for D2-branes
is isomorphic to the relative Jacobian variety of $X$, which is in turn 
isomorphic to $X$ itself since the elliptic fibration has only nodal 
and cuspidal fibers. Moreover, the number of moduli agrees with 
the number of marginal deformations computed at the Gepner point, 
supporting again the decoupling of K\"ahler moduli argued in \bdlr.\

Next, the states ${\bf n}_4\ldots{\bf n}_7$ have one unit of D6-charge.
${\bf n}_4$ is particularly interesting since the topological invariants 
\eqn\topE{
r(\eta)=1,\qquad \ch{1}{\eta}=0,\qquad 
\ch{2}{\eta}=0,\qquad \ch{3}{\eta}=-1}
identify the ideal sheaf of a single point $P$ on $X$. In the present 
conventions, this corresponds to a D6-D0 system which is very interesting. 
In flat space it is known that D6-D0 systems are repulsive and they 
break supersymmetry completely. On a curved manifold, the open string 
dynamics is harder to analyze, but in the large radius limit, the 
D6-D0 potential should approach continuously the flat space result. 
Therefore, we expect this configuration to be nonsupersymmetric and 
repulsive at sufficiently large radius. In particular, the repulsive
interaction prevents the occurrence of a bound state. This gives a clear 
example supersymmetric Gepner model state which decays in a 
nonsupersymmetric combination of D-branes in the large radius limit. 
Such phenomena have been predicted in \bdlr.\ This picture is 
especially interesting when interpreted from the mirror ${\hat X}$ point 
of view. As also mentioned in the introduction, we obtain an example 
of a phase transition of special lagrangian cycles as we move on the 
complex structure moduli space of ${\hat X}$. Regarding mirror 
symmetry as T-duality as in \SYZ,\ $X$ and ${\hat X}$ admit special 
lagrangian fibrations with dual $T^3$, ${\hat T}^3$ fibers. 
Then the D6 and D0-branes on $X$ are mapped to D3-branes wrapping 
the base $B$ of the fibration and respectively the ${\hat T}^3$ fiber. 
The previous argument shows that the homology class $B+{\hat T}^3$ 
should not support a special lagrangian cycle in a neighborhood 
of the large complex structure limit of ${\hat X}$. However, it should 
support such a cycle in a region of the moduli space of ${\hat X}$ 
which maps to a neighborhood of the Gepner point of $X$ under the mirror 
map. This predicts transitions of the type discussed in \DJ\ in 
a concrete compact model. 

The remaining states ${\bf n}_{5,6,7}$ correspond to various D6, D4
and D2 combinations whose existence as bound states is an open problem. 
The topological charges can be easily computed as above, but we will 
not pursue this here. 

Finally, the last two states ${\bf n}_{8,9}$ represent configurations 
with a D4-brane wrapped on the section $E$ of the elliptic fibration 
and a D2-brane wrapped on an elliptic fiber $Y$. For ${\bf n}_9$, the 
D4-brane carries a magnetic flux on a hyperplane in ${\bf P}^2$ 
which induces a lower D2-brane charge. Here, the dynamics is very similar 
to that of the D6-D0 system (in fact they can be related by a T-duality 
on the elliptic fiber). As $X$ approaches the large radius limit, 
the brane systems approach configurations of transverse D4-D2 
branes in flat space. These are nonsupersymmetric and repulsive, 
therefore they cannot form bound states. Hence, we find more examples 
of Gepner model states which decay into non-BPS configurations 
in the large volume limit. They should be interpreted in terms 
of phase transitions of special-lagrangian cycles on ${\hat X}$, 
as before.

The other series of Gepner model boundary states can be analyzed 
similarly, resulting in various brane configurations. We will not 
pursue this systematically here, but we would like to emphasize 
two other states which have not appeared in the previous cases. 
Namely, in the series $L=(2100)$ (which is identical to $L=(7000)$), 
one finds the charge vectors 
${\bf n}_1=(0,0,1,-1,-3,-1)$ and ${\bf n}_2=(0,0,1,-1,-1,0)$ 
which correspond to D4-branes wrapping a vertical holomorphic 
four-cycle $i:D\hookrightarrow X$ in the class $S$. Note that 
the surfaces in this class 
are generically smooth elliptic fibrations over a rational curve in the 
class $l$. The associated K theory classes are 
determined by the torsion sheaves 
$i_{\star}\left(J_{P+Q}\otimes \co_D(-h-l)\right)$ 
and $i_{\star}\left(J_{P+Q}\otimes \co_D(h)\right)$, where 
$J_{P+Q}$ is the ideal sheaf of two (possible coincident) points on $D$.
Therefore, in both cases, we obtain a single D4-brane wrapped on $D$ 
with two units of D0-charge which correspond to D0-branes located at the
points $P, Q$. The D4-branes also carry magnetic flux inducing lower 
D2-brane charges on curves in the class $-h-l$ and respectively $h$.
The existence of bound states with these charges is an open problem, 
but the large radius limit analysis suggests that they might exist
since the D4-D0-systems are supersymmetric in flat space. 
In fact, the bound states would have to be marginal, which makes 
the existence problem very subtle. 

\centerline{\bf Acknowledgments}

We are very grateful to Michael Douglas for suggesting the problem 
and for collaboration at an early stage of this work. We would 
also like to thank Ilka Brunner, Jaume Gomis, Paul Horja, Albrecht 
Klemm, Greg Moore and Ronen Plesser for valuable discussions. 
The work of D.-E. D. has
been supported by DOE grant DE-FG02-90ER40542 and the work of C. R. has 
been supported by DOE grant DE-FG02-96ER40559.

\listrefs
\end